# Prediction of a new ground state of superhard compound $B_6O$ at ambient conditions

Huafeng Dong[1,4], Artem R. Oganov[2,3,4,5,*], Qinggao Wang[3,6], Sheng-Nan Wang[4], Zhenhai Wang[4,7], Jin Zhang[4], M. Mahdi Davari Esfahani[4], Xiang-Feng Zhou[4,8], Fugen Wu[1], and Qiang Zhu[4]

[1] School of Physics and Optoelectronic Engineering, Guangdong University of Technology, Guangzhou 510006, China
[2] Skolkovo Institute of Science and Technology, Skolkovo Innovation Center, 3 Nobel St., Moscow 143026, Russia.
[3] Moscow Institute of Physics and Technology, 9 Institutskiy Lane, Dolgoprudny city, Moscow Region 141700, Russian Federation
[4] Department of Geosciences and Center for Materials by Design, Institute for Advanced Computational Science, State University of New York, Stony Brook, NY 11794, USA
[5] Northwestern Polytechnical University, Xi'an 710072, China
[6] Department of Physics and Electrical Engineering, Anyang Normal University, Anyang, Henan Province 455000, China
[7] Peter Grünberg Research Center, Nanjing University of Posts and Telecommunications, Nanjing 210003, China
[8] School of Physics and Key Laboratory of Weak-Light Nonlinear Photonics, Nankai University, Tianjin 300071, China

*Corresponding author. E-mail: artem.oganov@stonybrook.edu (A.R.O.)

**Boron suboxide $B_6O$, the hardest known oxide, has an $R\bar{3}m$ crystal structure (α-$B_6O$) that can be described as an oxygen-intercalated structure of α-boron, or, equivalently, as a cubic close packing of $B_{12}$ icosahedra with two oxygen atoms occupying all octahedral voids in it. Here we show a new ground state of this compound at ambient conditions, *Cmcm*-$B_6O$ (β-$B_6O$), which in all quantum-mechanical treatments that we tested (GGA, LDA, and hybrid functional HSE06) comes out to be slightly but consistently more stable. Increasing pressure and temperature further stabilize it with respect to the known α-$B_6O$ structure. β-$B_6O$ also has a slightly higher hardness and may be synthesized using different experimental protocols. We suggest that β-$B_6O$ is present in mixture with α-$B_6O$, and its presence accounts for previously unexplained bands in the experimental Raman spectrum.**

Ultrahard materials are used in many applications, from cutting, grinding and drilling tools to wear-resistant coatings [1]. However, most ultrahard materials [2], such as diamond [3] and cubic-BN [4], are synthesized at high pressure, which makes them expensive, but some (boron allotropes, $B_6O$, $B_4C$) are thermodynamically stable already at ambient conditions. The hardness of α-$B_6O$ [5] was reported to be in the range 30-45 GPa [6, 7], making it the hardest known oxide [7].

Objects with icosahedral symmetry ($I_h$) bear a special fascination; natural examples are rare, because of incompatibility of fivefold symmetry with crystalline periodicity. The discovery of multiply-twinned particles $B_6O$, an icosahedral packing of $B_{12}$ icosahedra with $I_h$ symmetry, had aroused widespread interest [5]. Here we report the prediction of a new phase of $B_6O$, with space group *Cmcm*, which we name β-$B_6O$. This structure is energetically almost degenerate with α-$B_6O$ (but slightly more stable than it), is predicted to have a higher hardness, and actually corresponds to twinned α-$B_6O$ structure.

**Discovery of β-$B_6O$ at ambient conditions**

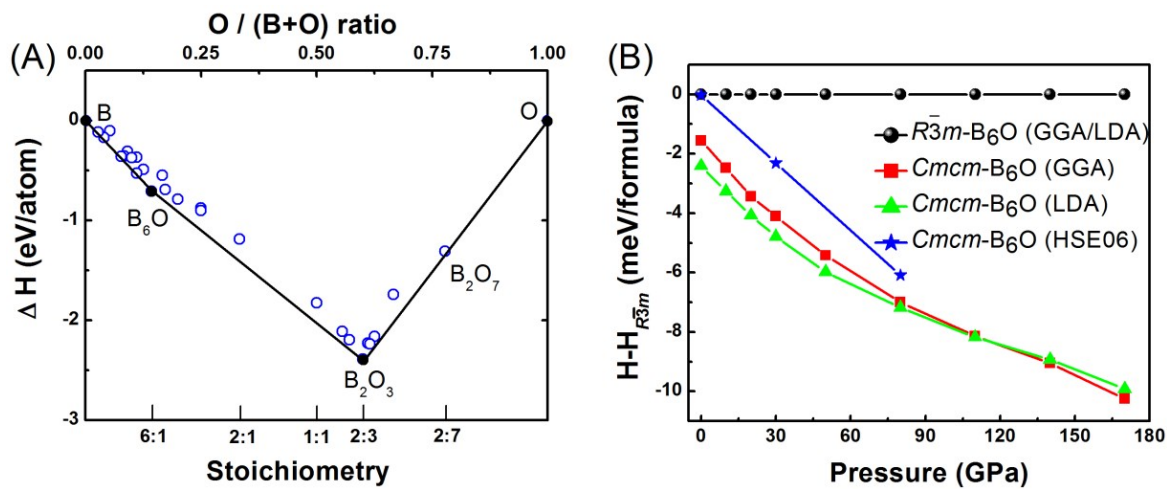

Fig. 1. **Stability of *Cmcm*-$B_6O$.** (A) Convex hull of the B-O system at ambient pressure. The solid (hollow) points represent the stable (metastable) structures. (B) Enthalpy difference between β-$B_6O$ and α-$B_6O$, including zero-point energy.

Our calculations expectedly find $B_2O_3$ and $B_6O$ to be the only stable compounds in the B-O system. Interestingly, there are also several compounds very close to stability - $B_2O_7$ (this is a 2D-form of $B_2O_3$ intercalated with oxygen molecules) and oxygen-deficient versions of $B_6O$ with $B_6O$-like structures and compositions between B and $B_6O$. To our surprise, $Cmcm$-$B_6O$ (β-$B_6O$, see Table 1 for structural parameters), instead of the well-known $R\bar{3}m$-$B_6O$ (α-$B_6O$) [5, 8-10], turned out to be the most stable structure at ambient pressure, as shown in Fig. 1. Transmission electron microscope result confirms its existence [11] and phonon calculations also confirmed its dynamical stability. Structural parameters and some of the physical properties of β-$B_6O$ are shown in Table 1, in comparison with α-$B_6O$ and two related forms of pure boron.

Table 1. **Structural parameters, hardness and band gap** of α-B, $Cmcm$-B, α-$B_6O$, and $Cmcm$-$B_6O$ phases.

| Phase | α-B | $Cmcm$-B | α-$B_6O$ | $Cmcm$-$B_6O$ |
|---|---|---|---|---|
| Space group | $R\bar{3}m$ | $Cmcm$ | $R\bar{3}m$ | $Cmcm$ |
| V, Å³/atom | 7.248 | 7.262 | 7.387 | 7.384 |
| Cell parameters | $a=b=c=5.050$ Å, $\alpha=58.04°$ | $a=4.883$ Å, $b=8.852$ Å, $c=8.064$ Å, $\alpha=\beta=\gamma=90°$ | $a=b=c=5.153$ Å, $\alpha=63.10°$ | $a=5.393$ Å, $b=8.777$ Å, $c=8.736$ Å, $\alpha=\beta=\gamma=90°$ |
| Atomic coordinates | B1(0.654,0.010, 0.010) B2(0.630,0.221, 0.221) | B1(0.000,0.236,0.568) B2(0.500,0.937,0.576) B3(0.823,0.167,0.750) B4(0.797,0.831,0.639) B5(0.682,0.995,0.750) | B1(0.998,0.998, 0.667) B2(0.676,0.201, 0.201) O(0.622,0.622, 0.622) | B1(0.000,0.756, 0.588) B2(0.000,0.549, 0.584) B3(0.165,0.824, 0.750) B4(0.238,0.155, 0.649) B5(0.334,0.987, 0.750) O(0.000,0.840, 0.439) |
| $Hv_{(Chen)}$, GPa | 39 | 35 | 38 | 39 |
| $Hv_{(Lyakhov)}$, GPa | 33.0 | 32.7 | 31.6 | 31.7 |
| DFT band gap, eV | 1.457 | 1.772 | 1.854 | 1.805 |

In order to further compare the stability of β-$B_6O$ and α-$B_6O$, we calculated their enthalpies as a function of pressure, as shown in Fig. 1B. We found that the enthalpy of β-$B_6O$ is lower than that of α-B$_6$O at ambient pressure, but the energy difference is only about 1.8 meV/formula within the GGA (and almost degenerate within the HSE06 hybrid functional). As pressure increases, β-B$_6$O becomes progressively more favorable than α-B$_6$O, indicating that β-B$_6$O might be more easily synthesized under high pressure. The energy of the two structures is so close that it makes us think: will the two structures coexist? what is their relationship? how to synthesize β-B$_6$O? In order to answer these questions, we make a detailed comparison of their structure, Raman spectra and phonon densities of states (PDOS) in the following.

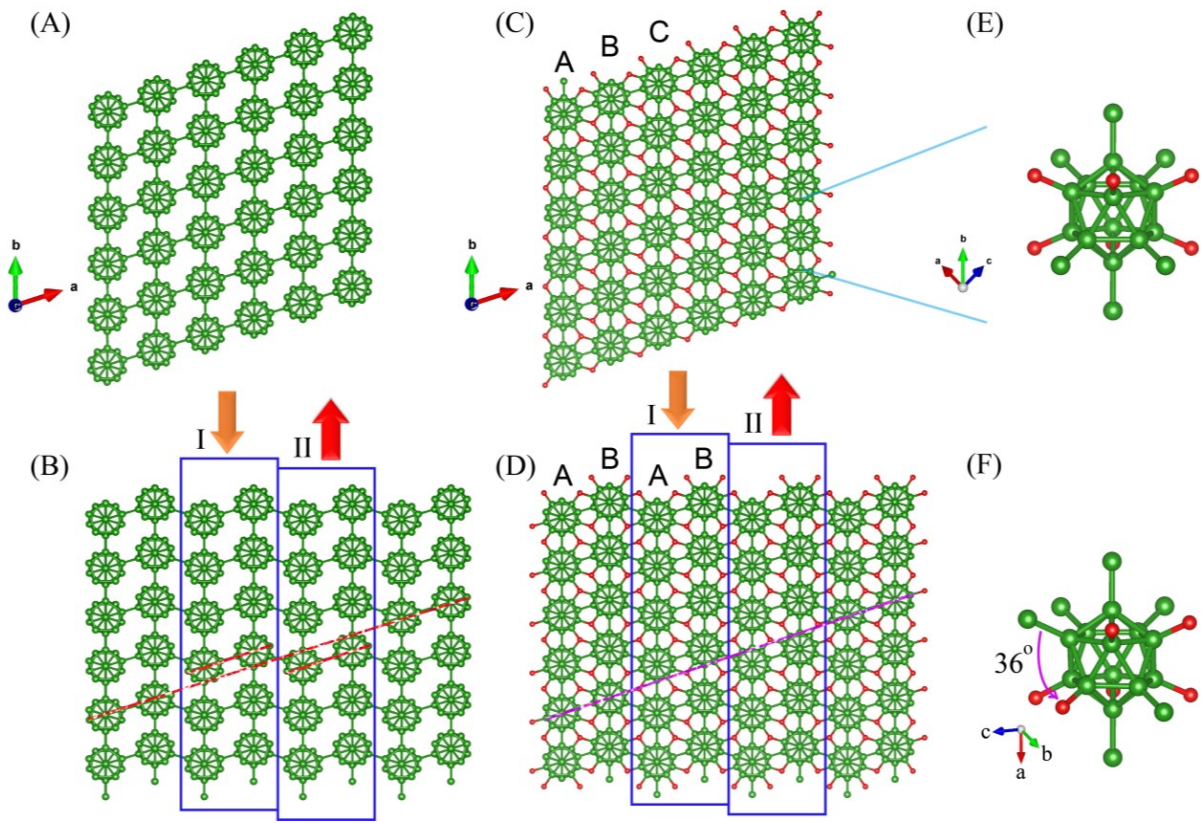

Fig. 2. **Crystal structures** of (A) α-B, (B) *Cmcm*-B, (C) α-B$_6$O, (D) *Cmcm*-B$_6$O, and their local structures, (E) and (F) B$_{12}$ icosahedra. Green (large) and red (small) spheres denote B and O atoms, respectively.

**Comparision of structure of α-B$_6$O and β-B$_6$O**

β-B$_6$O structure has hexagonal close packing of B$_{12}$ icosahedra (ABAB… stacking), while α-B$_6$O is based on the cubic close packing (ABCABC… stacking) of B$_{12}$ icosahedra, as shown in Fig. 2(D and C). As is the case of hcp and fcc metals, twinning of α-B$_6$O can produce local β-B$_6$O stackings. It may also be possible to obtain β-B$_6$O-like stacking faults by deformation of α-B$_6$O, through plane slips. Most properties of these two phases are very similar: e.g. predicted volume per formula ($V$(α-B$_6$O) = 51.71 Å$^3$/formula; $V$(*Cmcm*-B$_6$O) = 51.69 Å$^3$/formula), hardness ($Hv$ (α-B$_6$O) = 38 GPa; $Hv$ (*Cmcm*-B$_6$O) = 39 GPa), DFT band gaps (α-B$_6$O has a 1.85 eV direct band gap, while *Cmcm*-B$_6$O has a 1.81 eV indirect band gap).

Another interesting aspect is that if we remove the oxygen atoms from α-B$_6$O and *Cmcm*-B$_6$O, they turn into α-B [12] and *Cmcm*-B, respectively (Fig. 2 (A and B)). α-B and β-B are the stable structures of boron at low pressure [13], while *Cmcm*-B is a newly predicted, and energetically nearly degenerate with them, structure [14, 15]. As shown in Fig. 2B, displacements of layers I and II can transform this structure into α-B, and vice versa; *Cmcm*-B$_6$O and α-B$_6$O have a similar relationship (Fig. 2D). Furthermore, we found that the conversion of *Cmcm*-B$_6$O and α-B$_6$O will cause a deflection of B-O bond by 36°, as shown in their local structure (Fig. 2 (E and F)).

However, it should be pointed out that it is not easy to change the stacking in covalent systems. However, examples are known – lonsdaleite (metastable "hexagonal diamond") is formed in shocked cubic diamond. To obtain multiple polytypes, methods like physical vapor transport (PVT), also known as seeded sublimation growth, can be suitable: e.g., different polytypes of SiC were obtained using the PVT method [16].

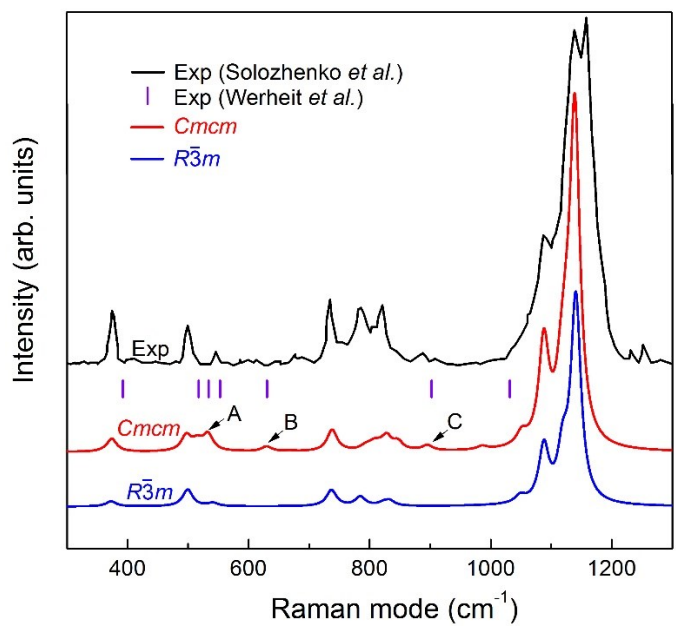

Fig. 3. **Raman spectra of $B_6O$:** experimental spectrum of Solozhenko [8], Raman mode frequencies experimentally observed by Werheit and Kuhlmann [17], and our calculated Raman spectra of *Cmcm*-$B_6O$ and α-$B_6O$.

**Comparision of Raman spectra**

As mentioned above, β-$B_6O$ and α-$B_6O$ are energetically almost degenerate at zero pressure, structurally related and can coexist. To test this hypothesis, we calculated their Raman spectra and compared with the experimental results [8, 9, 17]. In Fig. 3, the topmost curve is the Raman spectrum reported by Solozhenko [8]; below it are the Raman frequencies reported by Werheit and Kuhlmann [17] and marked by vertical bars (|). The two curves below are our Raman spectra of β-$B_6O$ and α-$B_6O$. As one could expect, the Raman spectra of β-$B_6O$ and α-$B_6O$ are similar and match perfectly the experimental spectra. For example, the Raman modes at 499, 541, 737, 785, 833, 1088, 1119, 1141 cm$^{-1}$ are consistent with Solozhenko's results [8]. However, there are four Raman modes which are unique for β-$B_6O$. The first two are computed to be at 516 and 533 cm$^{-1}$ (marked by letter A in Fig. 3), and Werheit and Kuhlmann indeed observed these two modes at 519 and 534 cm$^{-1}$. The third mode is predicted to be at 630 cm$^{-1}$ (marked with letter B), and Werheit and Kuhlmann have indeed observed a Raman-active phonon at 627 cm$^{-1}$. We note that α-B$_6$O does not have Raman-active modes between 570 and 700 cm$^{-1}$, thus the one observed by Werheit and Kuhlmann at 627 cm$^{-1}$ cannot come from α-B$_6$O, but β-B$_6$O. The fourth mode has the theoretical Raman frequency of 896 cm$^{-1}$ (marked with letter C), while α-B$_6$O has no Raman-active modes between 850 and 1000 cm$^{-1}$. Thus, this mode is also unique to the β-B$_6$O structure, and again seen in experiments: Solozhenko observed Raman active phonon at 889 cm$^{-1}$, and Werheit and Kuhlmann observed a Raman-active phonon at 902 (909) cm$^{-1}$. Moreover, Wang *et al.* also observed similar Raman spectra in their B$_6$O samples (see Fig. 1 in Ref. [9]). This analysis clearly shows that experimental samples contain β-B$_6$O.

**Comparision of PDOSs and Gibbs free energy**

Comparing phonon densities of states (PDOSs) of α-B$_6$O and β-B$_6$O at ambient pressure (Fig. 4), we once again see a great degree of similarity. In order to further confirm the stability of β-B$_6$O, we have calculated the Gibbs free energy of β-B$_6$O and α-B$_6$O as a function of temperature, shown in Fig. S3. We conclude that β-B$_6$O remains more stable than α-B$_6$O also when temperature is taken into account – and is even slightly stabilized by thermal effects.

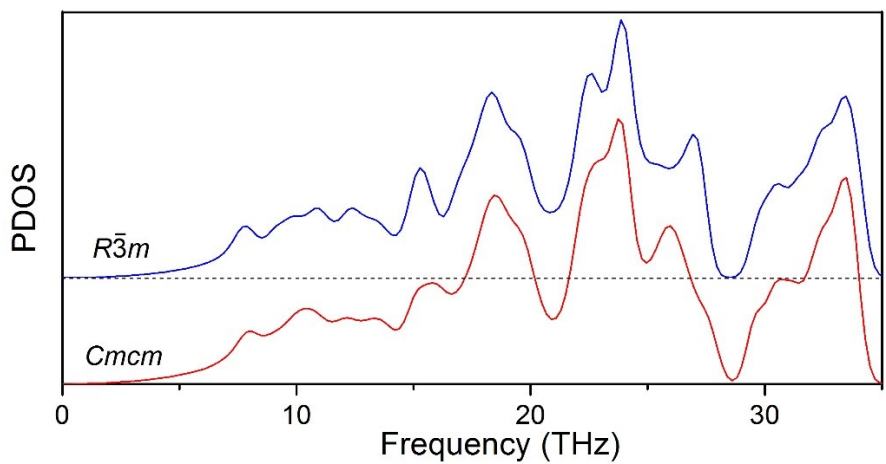

Fig. 4. **Phonon density of states** of α-B$_6$O phase and *Cmcm*-B$_6$O phase at ambient pressure. For clarity, the PDOS of α-B$_6$O was shifted.

Having found that β-$B_6O$ should be more stable than α-$B_6O$, and demonstrated that the two phases actually coexist in experimental samples, we ask a question: why is the synthetic compound mostly α-$B_6O$, instead of the more stable β-$B_6O$? While there can be no definitive answer at this point, we suggest that this may be because of the use of α-rhombohedral-B ($R\bar{3}m$) [7] or β-rhombohedral-B ($R\bar{3}m$) [18] as a starting material (together with $B_2O_3$) for synthesis of rhombohedral-$B_6O$. To obtain β-orthorhombic-$B_6O$, one would need to crystallize it *from the melt* (preferably at high pressure, to increase its thermodynamic advantage over α-$B_6O$), or use other phases of boron as precursors. Using *Cmcm*-B would be ideal, but this phase (just 11 meV/atom higher in energy than α-B [15]) remains hypothetical, though likely to be eventually synthesized. Alternatively, the PVT method [16] could be used.

In summary, to our big surprise, *ab initio* structure prediction calculations discovered a new ground state for the widely studied superhard compound $B_6O$ – our predicted β-$B_6O$ is more stable that experimentally established α-$B_6O$. The two phases are polytypes and have nearly the same densities (β-$B_6O$ is slightly denser), energies (slightly lower for β-$B_6O$), band gaps (slightly smaller and indirect, rather than direct, for β-$B_6O$), hardnesses (β-$B_6O$ is slightly harder) and phonon densities of states, but have important differences in Raman spectra. By comparing calculated and experimental Raman spectra, we demonstrated that the experimental samples are actually a mixture of α-$B_6O$ and β-$B_6O$. The discovery of β-$B_6O$ opens up new possibilities, in view of its greater stability and hardness and indirect band gap. Our findings also indicate possibilities of tuning the properties of $B_6O$ by obtaining phase-pure samples (probably not obtained to date), and the possibility of metastable oxygen-deficient compounds based on α-$B_6O$ or β-$B_6O$ - these can be obtained at high temperatures (where disordered oxygen vacancies will stabilize the structure) and low chemical potentials of oxygen.

## Methods

We used the *ab initio* evolutionary algorithm USPEX [19-22] to search for thermodynamically stable B-O compounds and their structures at ambient pressure. This methodology has shown its predictive power in many studies (e.g., [13, 22-24]). All structures were relaxed; structure relaxations and total energy calculations were done using density functional theory (DFT) within the generalized gradient approximation (GGA) [25] as implemented in the VASP code [26], with the projector-augmented wave method [27]. We used plane-wave kinetic energy cutoff of 600 eV, and sampled the Brillouin zone with uniform Γ-centered meshes of is $2\pi*0.07$ Å$^{-1}$ resolution within structure search, and $2\pi*0.04$ Å$^{-1}$ for subsequent highly precise relaxations and properties calculations. In order to confirm the relative stability of α-$B_6O$ and β-$B_6O$, we used local density approximation (LDA) [28] and HSE06 hybrid functional [29]. Phonon spectra is computed by PHONOPY [30] and VASP, and Raman spectra were calculated using the Fonari-Stauffer method [31]. Hardness was calculated with Chen model [32] and Lyakhov-Oganov model [33].

## Acknowledgments

**Funding:** H. F. D. gratefully acknowledges financial support from the National Natural Science Foundation of China (Grant No. 11547174). A. R. O. thanks the Government of Russian Federation (grant No.14.A12.31.0003), and Foreign Talents Introduction and Academic Exchange Program (No. B08040). Calculations were carried out in part at the Center for Functional Nanomaterials, Brookhaven National Laboratory, which is supported by the U.S. Department of Energy, Office of Basic Energy Sciences, under Contract No. DE-AC02-98CH10886. This work used the Extreme Science and Engineering Discovery Environment (XSEDE), which is supported by National Science Foundation grant number ACI-1053575.

**Figures legends**

Fig. 1. **Stability of *Cmcm*-B$_6$O.** (A) Convex hull of the B-O system at ambient pressure. The solid (hollow) points represent the stable (metastable) structures. (B) Enthalpy difference between β-B$_6$O and α-B$_6$O, including zero-point energy.

Fig. 2. **Crystal structures** of (A) α-B, (B) *Cmcm*-B, (C) α-$B_6O$, (D) *Cmcm*-$B_6O$, and their local structures, (E) and (F) $B_{12}$ icosahedra. Green (large) and red (small) spheres denote B and O atoms, respectively.

Fig. 3. **Raman spectra of $B_6O$:** experimental spectrum of Solozhenko [8], Raman mode frequencies experimentally observed by Werheit and Kuhlmann [17], and our calculated Raman spectra of *Cmcm*-$B_6O$ and α-$B_6O$.

Fig. 4. **Phonon density of states** of α-$B_6O$ phase and *Cmcm*-$B_6O$ phase at ambient pressure. For clarity, the PDOS of α-$B_6O$ was shifted.